\documentstyle[11pt,axodraw]{article}
\input psfig.tex 
\def\MEt{\mbox{$E\kern-0.57em\raise0.19ex\hbox{/},{T}$}\ }
\def\MEtp{\mbox{$E\kern-0.57em\raise0.19ex\hbox{/}_{T}$}}
\def\D0{\mbox{D\O }}
\begin{document}
\begin{center}
{\Large\bf Limits on Exotic Quarks in the
SU(3)$\times$U(1) 
Extension \\ of the Standard Model from SUSY Search Data}

\vspace{0.5in}
{\large Prashanta Das$^a$, Pankaj Jain$^a$ and 
Douglas W. McKay$^b$}

\bigskip
$^a$Physics Department,
I.I.T. Kanpur,
India 208016\\

\bigskip
$^b$Department of Physics \& Astronomy\\
University of Kansas,\\
Lawrence, KS 66045,
USA
\end{center}

\noindent
{\large\bf Abstract:} 
We study the $p\bar p$ production and decay of 
exotic quarks that
are predicted by the SU(3)$\times$U(1) extension of the Standard Model.
We show that recent experimental searches for SUSY particles
at the Tevatron limit the mass of these quarks to be above 250 GeV.
Run II will extend the reach to 320 GeV. This is one example of 
SUSY search signatures that apply directly to another, quite different,  new
physics model. 

\vfill
\noindent
pdas@iitk.ac.in\\
pkjain@iitk.ac.in\\
mckay@kuark.phsx.ukans.edu\\

\newpage
Among the``beyond the standard" models that offer plausible motivation and 
some attrative features, the ``minimal" $SU(3)_{Color} \times SU(3)_L 
\times U(1) (3-3-1)$ model \cite{Pisano,Foot,Frampton92}  
distinguishes itself with a relativley low,
$O(1\ {\rm TeV})$ 
scale of new physics and some spectacular gauge boson and heavy
quark decay patterns. Phenomenological studies \cite{Ng,
Frampton94,Dutta,Sasaki} to date have put the
emphasis on searches for, and mass limits on, signals of the new gauge
bosons at $e^+e^-$ and $pp$ colliders. In this note we focus instead on
limits that $p\bar p$ collisions at the Tevatron have set or will set on
the new heavy quarks.
 
The 3-3-1 models  have the special twist that anomaly cancellation
requires that there be  
three families of fermions, which is related to
the fact that the color group is $SU(3)$. The new quarks (new compared to
the Standard Model) are the third members of the fundamental
representations of $SU(3)_L$ along with their right--handed $SU(3)_L$
singlet counterparts. The assignment of representations is not unique, and
we choose the lepton $\underline{3^*}$ triplet to be

 \[
 \left( \begin{array}{c}
 e \\ \nu_e\\ e^c\\ \end{array}\right)_L \quad ,\quad
 \left( \begin{array}{c} 
 \mu \\ \nu_\mu\\ \mu^c\\ \end{array}\right)_L \quad ,\quad
 \left( \begin{array}{c}
 \tau \\ \nu_\tau\\ \tau^c\\ \end{array} \right)_L\quad ,
 \]
 with $U(1)$ charge, or ``strong" hypercharge, $Y_S=0$. The ``strong" refers 
to the fact that the physical vector boson that is predominantly the
$U(1)$ gauge boson has a large coupling to matter fields. Our definition
of $Y$ is: $Y\equiv \sqrt{3}\lambda_8-Y_S {\bf 1}_{3\times 3}$, were
$Q=T_3+{Y\over 2}$ as usual. The question of the neutrino masses does not
concern us here, so we do not consider the possibility of singlet,
right-handed neutrinos. The three quark families consist in $SU(3)_L$
triplet and singlet choices as follows: 
 
 \[ (1) \left( \begin{array}{l}
         u \\ d\\ D \\ \end{array}
         \right)_L \underline{3}, {2\over 3};\hskip .25in
         u_R\ \ \underline{1}, -{4\over 3}; \hskip .25in
         d_R\ \ \underline{1}, {2\over 3};\hskip .25in
         D_R\ \ \underline{1}, {8\over 3}\hskip .5in
 \]
 \[ (2) \left( \begin{array}{l}
         c \\ s\\ S \\ \end{array}
         \right)_L \underline{3}, {2\over 3};\hskip .25in
         c_R\ \ \underline{1}, -{4\over 3};\hskip .25in
         s_R\ \ \underline{1}, {2\over 3};\hskip .25in
         S_R\ \ \underline{1}, {8\over 3}\hskip .5in
 \]
  
 \[ (3) \left( \begin{array}{l}
         b \\ t \\ T \\ \end{array}
         \right)_L \underline{3^*}, -{4\over 3};\hskip .25in 
	b_R\ \ \underline{1},
         {2\over 3}; \hskip .25in
         t_R\ \ \underline{1}, -{4\over 3};\hskip .25in 
	T_R\ \ \underline{1}, -{10\over 3}
 \]
 
\noindent 
and it is understood that the leptons are $SU(3)_{color}$ singlets 
and all of the quarks are in triplet representations.
 
Our representation of the covariant derivative for $SU(3)_L \times U(1)$
interactions with matter fields in the fundamental representation is
 
 $$D_\mu=\partial_\mu-ig T^a W^a_\mu + i {g_S\over 2} Y_S V_\mu ,
 $$
 
 \noindent where
 $T^a={\lambda^a\over 2}$ in terms of the standard Gell--Mann Matrices
 $\lambda^a$, and a $3\times 3$ unit matrix is understood in the $V_\mu$, 
$U(1)$  gauge field, interaction. The gauge field matrix in more detail
reads
 
 \[ T^aW^a = {1\over \sqrt{2}}\left(\begin{array}{lllll}
                 {W^3\over \sqrt 2}+{W^8\over \sqrt{6}} & & W^+ & & Y^{++} \\
                 W^- & & {-W^3\over \sqrt 2}+{W^8\over \sqrt{6}} & & Y^+ \\
                 Y^{--}& & Y^- & & {-\sqrt 2\over \sqrt{3}}W^8 \\ 
   \end{array} \right)
 \]
 
 The $W^\pm$ are identified with the usual $W$-bosons of the standard model,
 while the $Y$'s, called dileptons, carry lepton number, $| L | =2$, and
 their interactions give rise to $Y^{--}\rightarrow e^-e^-$ and $Y^-
\rightarrow e^- \nu_e$ decays, for example. The heavy quarks $D,S,T$,
also with $| L | =2$, can decay through real or virtual $Y$ emission to
produce dilepton final states with a $| \Delta L| = 2$ decay signature.
These are the ``spectacular" decay patterns referred to above.
 
 Unless some special radiative mass generation effects are invoked, the 
minimal Higgs structure contains three $\underline{3}$'s and one
$\underline{6}$ of $SU(3)$ \cite{Foot}. 
In this minimal Higgs version of $3-3-1$
models, the limited number of vacuum expectation values leads to the mass
relation
\begin{equation}
M_Y/M_{Z'}=[3g^2/(4g^2 + 12g^2_S)]^{1/2}\ .
\end{equation}
Additional higgs multiplets allow one to relax this constraint. In addition,
requiring that the $SU(2)_L\times U(1)$ couplings match the $SU(3)_L\times
U_Y(1)$ couplings at the $SU(3)_L$ breaking
 scale $M_Y$ leads to the relationship 
\begin{equation}
g^2_S/g^2|_{M_Y}= \sin^2\theta_W/(1-4
 \sin^2\theta_W)|_{M_Y}\ .
\end{equation}
This expression shows the ``strong $g_S$" 
character of the model, since $4 \sin^2\theta_W=0.92$ at the $Z$ mass and
it grows as the
 scale $M_Y$ is increased \cite{Jain}. 
Since the Higgs fields play a negligible role in the
 production and decay of the $D, S$ and $T$ quark fields, we do not review 
their properties here.

\bigskip
\noindent 
{\it Analysis of $p\bar p\rightarrow Q\bar Q$ at 2 TeV}.
 
 We are specifically  addressing the bounds on new 3-3-1 quark properties that
 the Tevatron can set, so we assume that $M_Q<M_Y$ since previous studies 
\cite{Frampton94} have
 set bounds $>300$ GeV on $M_Y$ from $\mu - $ decay  
data.\footnote{Neutrino-oscillation ``appearance'' 
experiments set comparable limits of 
$> 340$ GeV \cite{Johnson98}.} 
Heavy quark masses larger than this will exceed the 
Tevatron's discovery reach even in run II, as we show below.
For this study we do not require the minimal model relation, Eq. (1),
between $M_Y$ and $M_{Z'}$. We fix $M_Y=300$ GeV (the results are 
insensitive to $M_Y$) and $g_S(M_Y)$ and allow $M_{Z'}$ to range upward
from one TeV. An expanded Higgs sector that introduces another vacuum
expectation value not fixed by $SU(2)_L$ breaking allows $M_{Z'}$ and
$M_Y$ to be chosen independently \cite{Dutta}.

 The parton level Feynman diagram relevant to the $Z'$ mediated production of
exotic quark 
 plus antiquark in hadron colliders is shown in Fig. 1, 
 while the
 semi-leptonic decay diagram of a heavy quark is shown in Fig. 2. The QCD
diagrams are the standard ones for heavy quark plus antiquark production. 
We use PYTHIA \cite{pythia} to compute the contribution to the $Q\bar Q$
 cross sections for both of
these channels.
The neutral
 $Z'$ spin-one boson is mixed with an 
angle $|\theta |< 5\times 10^{-3}$ with the
 physical $Z$, so we do not include this tiny effect in the subsequent
 discussion. The relevant couplings beyond QCD then read:
\begin{figure}[htb]
\vspace*{-15ex}
\begin{tabbing}
\begin{picture}(200,105)(-100,-0)
\ArrowLine(0,0)(50,0)
       \Text(25,+10)[c]{$p^{+}$}   
%%% \Vertex(50,0){3}
\BCirc(50,0) {9}
\ArrowLine(50,0)(135,0)
\ArrowLine(50,+4)(135,+4)
\ArrowLine(50,0)(100,-50)
       \Text(70,-25)[r]{$u,d$}
\ArrowLine(0,-100)(50,-100)
       \Text(25,-110)[c]{$p^{-}$}
%%% \Vertex(50,-100){3}
\BCirc(50,-100) {9}
\ArrowLine(50,-100)(135,-100)
\ArrowLine(50,-104)(135,-104)
\ArrowLine(50,-100)(100,-50)
       \Text(60,-75)[c]{$\overline {u}, \overline {d}$}  
%%% \Vertex(100,-50){3}
\Photon(100,-50)(150,-50){5}{4}
       \Text(140,-65)[r]{$Z^{\prime}$}

\ArrowLine(150,-50)(200,-25)
       \Text(175,-25)[l]{$\overline {D}$}
\ArrowLine(150,-50)(200,-75)
       \Text(175,-75)[l] {$D$}
\end{picture}
\end{tabbing}
\vspace {1.5in}

 \caption {\em $D  \overline {D}$ production from $p\overline {p}$
collision via
 $Z^{\prime}$}
\end{figure}

\vskip 0.2in 
\begin{equation}
 {\cal L}_{f\bar fZ'}=\sum\limits_{f=u,d,D,t}{g\delta\over \cos\theta_W}
\bar f\gamma_\mu (a_f + b_f
 \gamma_5)f  Z^{'\mu}
\end{equation} 
 and, with $C$ the charge conjugation operator, 
\begin{equation}
\begin{array}{lll} 
 {\cal L}_{f_1\bar f_2 Y}&=&{-g\over
 2\sqrt{2}}\sum\limits_{\ell=e,\mu,\tau}Y^{--}_\mu \bar\ell\gamma^\mu\gamma_5
 C\bar\ell^T +{g\over 2\sqrt{2}}Y_\mu^{--}\bar D \gamma^\mu (1-\gamma_5)u 
+h.c  \nonumber\\ & +& {g\over 2\sqrt{2}}\sum\limits_{\ell=e,\mu,\tau}Y^{-}_\mu 
\bar\nu_\ell 
\gamma^\mu(1-\gamma_5)C\bar\ell^T 
+ {g\over 2\sqrt{2}}Y_\mu^{-}\bar D
\gamma^\mu (1-\gamma_5)d +h.c, 
\end{array}
\end{equation}
 for the production and decay vertices. We assume for illustration that the 
$D$ quark is the lightest of the new quarks in the model and concentrate
our attention on its production and decay. Our bound on $M_D$ will be the
least lower bound in the sense that 
including one or both of the other exotic quarks to be
active or choosing $T$ as the lightest quark will yield larger lower bounds.
The factors $a_f$ and $b_f$ in
Eq. (3)
are summarized
in Table 1, and the coefficient $\delta$ is defined as  
$\delta\equiv (1-\sin^2\theta_W)/[3(1-4 
\sin^2\theta_W)]^{1/2}$.
 
 \begin{center}
 \begin{tabular}{|c|ccccccccc|}
\hline
  q & u  &d  &D  &c &s&S&t&b&T  \\
\hline
   &   &  &  & &&&&&  \\
  $a_f$ & ${1\over 6}$ & $-{1\over 3}$ & $-{5\over 6}$ & ${1\over 6}$
& $-{1\over 3}$& $-{5\over 6}$& ${2\over 3}$& ${1\over 6}$& ${7\over 6}$\\
   &   &  &  & &&&&&  \\
\hline
   &   &  &  & &&&&&  \\
  $b_f$ & ${1\over 2}$ & $0$ & $-{1\over 2}$ & ${1\over 2}$
& $0$&  $-{1\over 2}$&$0$& $-{1\over 2}$& ${1\over 2}$\\
   &   &  &  & &&&&&  \\
\hline
 \end{tabular}
 \end{center}
 
 \begin{center}
 Table 1
 \end{center}
\begin{figure}[htb]
\vspace*{-13ex}
\begin{tabbing}
\begin{picture}(175,100)(0,-0)
\ArrowLine(0,0)(50,0)
       \Text(25,+10)[c]{$D(p)$}   
%%% \Vertex(50,0){3}
\ArrowLine(50,0)(100,50)
       \Text(75,35)[r]{$u(p^{\prime})$}
\DashLine(50,0)(100,-15){4}  
       \Text(75,-12)[r]{$Y^{--}$}
%%% \Vertex(100,-15){5}
\ArrowLine(100,-15)(175,-7)
       \Text(130,-4)[l]{$l^{-}(k^{\prime})$}
\ArrowLine(100,-15)(175,-45)
       \Text(120,-40)[l]{$l^{-}(k)$}
\ArrowLine(250,0)(300,0)
       \Text(275,+10)[c]{$ D(p)$}   
%%% \Vertex(200,0){3}
\ArrowLine(300,0)(350,50)
       \Text(325,35)[r]{$d(p^{\prime})$}
\DashLine(300,0)(350,-15){4}  
       \Text(325,-12)[r]{$Y^{-}$}
%%% \Vertex(100,-15){5}
\ArrowLine(350,-15)(425,-7)
       \Text(380,-4)[l]{$l^{-}(k^{\prime})$}
\ArrowLine(350,-15)(425,-45)
       \Text(370,-40)[l]{$\nu_l(k)$}
\end{picture}
\end{tabbing}
\vspace {0.5in}

 \caption {\em decay modes of $D$ quark}
\end{figure}
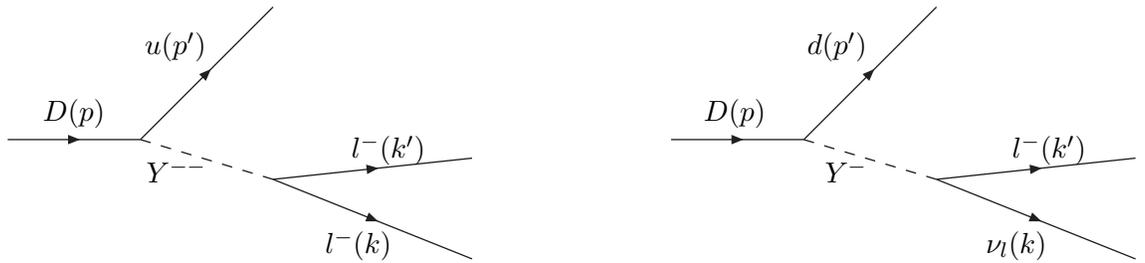

In Fig. 3 we show the contribution to the $p\bar p \rightarrow t\bar t+ X$
cross-section from QCD \cite{Berger,Laenen,
Catani} plus the $Z'$ graph in Fig. 1 as a function of $M_{Z'}$. We have
used a fixed coupling $\delta=1.89$ corresponding to a scale of
300 GeV, and the value $M_t=175$ GeV, in making this estimate. 
Consistent with our remarks above, we also 
assume that besides $D$ the remaining exotic quarks are too heavy to constitute a decay mode of $Z'$. Requiring that QCD cross-section plus 
the $Z'$ contribution, which adds incoherently to the QCD cross section, 
lie within the experimental $1-\sigma$ uncertainty \cite{cdf}, 
yields a lower bound on
$M_{Z'}$ between 750 GeV and 1100 GeV. 
Our bound from $t\bar t$ production is independent of the details of the
Higgs sector and, though weaker than the bound from the $Z'$ contribution
to $\Gamma_b$ in $Z^0$ decay \cite{Jain}, it may become quite stringent as
the experimental uncertainties in the $t\bar t$ cross-section are reduced. We
should remark that the $2\sigma$ lower bound of 1.7 TeV found in [8] prevents
us from suggesting that Fig. 3 shows evidence for $Z'$ in the mass range 750
GeV $\leq M_{Z'}\leq 1100$ GeV. 
 
Next we plot the cross section for $p\bar p\rightarrow D\bar D +X$ in Fig. 4 
as a function of $M_D$ for $M_{Z'} =1$ TeV and $M_{Z'}=\infty$; the latter
is the pure QCD case which is the same as the top quark case, and it sets a
minimum, or weakest, bound on the heavy quark mass $M_D$, as we discuss below.

The recently published results of  searches for SUSY particles  at the
Tevatron \cite{susy,abe}  
allow us to put 
limits on the exotic quarks  and illustrate that SUSY
searches can be applied with little cost to other new physics processes. 
These studies conducted by the \D0 and CDF detector groups 
use trilepton final states \cite{susy,abe} 
that could arise from the decay
of the charginos and
neutralinos  $\tilde \chi_1^\pm/\tilde \chi_2^{0}$. $\tilde \chi_1^\pm$
decays into a charged lepton, a neutrino and LSP and the $\tilde\chi_2^0$ decays
into two charged leptons plus an LSP. Simulating the same lepton isolation cuts
as used in these experiments we may apply their results
directly to a  search for 
the exotic 3-3-1 quarks, which
decay into a jet plus two like charged leptons or a jet plus a charged
lepton and a neutrino. The $D \bar D$ final state lepton products are $\ell
\ell \bar\ell$ or $\bar\ell\bar\ell \ell$, just as the SUSY case. 

The search for gauginos was conducted by considering the 
$eee$,$ee\mu$,$e\mu\mu$ and $\mu\mu\mu$ 
trilepton channels. The \D0 experiment [16] turns out to provide the more
strigent limit on the exotic quark mass, so we present some details of that
case here.         
For the $eee$ channel two different cuts were imposed in the experimental
search corresponding to two different triggers $e\MEtp$ and 
$2e\MEtp$. The first trigger $e\MEtp$ 
required that at least one electron is
detected with transverse energy $E_T$ greater than 20 GeV and the missing
energy $\MEtp$ is greater than 15 GeV. The second trigger 
$2 e\MEtp$ required detection of at least one electron with $E_T>12$
GeV and at least one more electron with $E_T>7 $ GeV and missing energy
$\MEtp>7$ GeV. 
The minimum lepton transverse energies  
$E_{T1}$,$E_{T2}$ and $E_{T3}$ (GeV) are required to be 
22,5,5 and 14,9,5 for the two triggers $e\MEtp$ and $2e\MEtp$ respectively. 
Similar cuts are placed for the
channels $ee\mu$,$e\mu\mu$.
 For case of three muons the corresponding 
values are 17,5,5 and 5,5,5 with the trigger
 $\mu$ and $\mu\mu$ respectively.

If $D\bar D$ is light enough to be produced at the Tevatron, it  
will also contribute to the cross-section for trilepton plus missing energy
signal used to search for charginos and neutralinos. 
We consider a specific channel
in which the trilepton final state consists of three electrons. 
Integrating
over the produced jets, there are six decay modes available to the
$D$ quark, namely $\ell_i\ell_i$ and $\ell_i\nu$ where $i=1,2,3$ 
corresponds to the three generations and $\ell_i$ represents the 
charged leptons. With the corresponding six 
decay modes of $\bar D$ we get a total of 36 decay channels for $D\bar D$.
Since either $D$ or $\bar D$ can decay into two electrons the total
for production of the final state $eee\nu$ is $1/18$ of the 
total cross section of $D\bar D$ production.  
We determine the number of such trilepton events expected from the $D\bar D$
final state putting the cuts as prescribed by the experimental search 
\cite{susy}. 

In brief, we take the PYTHIA generated events from both QCD and $Z'$ production
and decay of $D\bar D$ and require the lepton $E_T$ cuts described above and 
$\MEtp > 15$ GeV and $|\eta|<3.5$ for electrons and $|\eta|< 1.0$ for 
muons. A lepton isolation cut is applied to all events: if a nominal jet has
$P_T > 15$ GeV, then $R\equiv\sqrt{(\Delta\eta)^2+(\Delta\phi)^2}
>0.4$, is required where $\Delta\eta = |\eta_{\rm jet}-\eta_{\rm lepton}|$ and
$\Delta\phi = |\phi_{\rm jet} - \phi_{\rm lepton}|$. An efficiency 
factor of $1/4$ is applied in order to summarize the lepton 
identification efficiencies and electron tracking efficiency. The 
ratio of total efficiency to kinematic plus trigger efficiencies roughly
accounts for these experimental effects \cite{susy}.

For our purpose it suffices to consider the channels $eee$ and $\mu\mu\mu$.
We impose the appropriate cuts and efficiency factor as just 
described to estimate the number of $3l+\MEtp$ events expected in a
95 pb$^{-1}$ sample. Noting that the experiment showed no events after
the full cuts were applied and that the backgrounds are less than 0.5
events, we require that fewer than 0.5 events are obtained in the simulated
sample for each pair of $M_D,\ M_{Z'}$ values at the $M_Y$ value chosen. 
Overall the
$eee$ cuts are looser (the $\eta$ cuts are the determining factors), 
so the $M_D$ value for a given $M_Z$ is 
higher, than for the $\mu\mu\mu$ case. In Fig. 5 we show that resulting
exclusion boundary in $M_D-M_{Z'}$ space for $M_Y=300$ GeV. The region below
the curve is excluded, and the $M_{Z'}\rightarrow\infty$ limit gives the
(QCD determined) lowest bound $M_D>250$ GeV, which is our main result.   

\bigskip
\noindent
{\it Conclusions}

\medskip
The $SU(3)\times U(1)$ extension of the standard model has reasonable
motivation \cite{pires} 
and strong, easily testable predictions that have some features in common with
the predictions of supersymmetry. We applied the results of  recent 
\D0 and CDF searches for gauginos in $3l\MEtp$ final states to put a new, lower 
bound of 250 GeV/c$^2$ on the mass of exotic quarks in the 
$SU(3)\times U(1)$ model. The increased luminosity and 
energy at run II should be able to push this bound up to about 320 GeV.
 
The statement of our result can be turned around. Had \D0 obtained
a signal above background, it could have been interpreted as a signal
for neutralino and a chargino production or for exotic $SU(3)\times U(1)$ quark
production.  
Further study of the jet activity in the events would have been necessary to
discriminate between the two. 
We suggest that it would be useful to include a survey of such
alternatives to SUSY interpretations in the analysis 
of ``SUSY signal'' searches.

\bigskip
% GNUPLOT: LaTeX picture
\setlength{\unitlength}{0.240900pt}
\ifx\plotpoint\undefined\newsavebox{\plotpoint}\fi
\sbox{\plotpoint}{\rule[-0.200pt]{0.400pt}{0.400pt}}%
\begin{picture}(1500,900)(0,0)
\font\gnuplot=cmr10 at 10pt
\gnuplot
\sbox{\plotpoint}{\rule[-0.200pt]{0.400pt}{0.400pt}}%
\put(220.0,113.0){\rule[-0.200pt]{4.818pt}{0.400pt}}
\put(198,113){\makebox(0,0)[r]{2}}
\put(1416.0,113.0){\rule[-0.200pt]{4.818pt}{0.400pt}}
\put(220.0,240.0){\rule[-0.200pt]{4.818pt}{0.400pt}}
\put(198,240){\makebox(0,0)[r]{4}}
\put(1416.0,240.0){\rule[-0.200pt]{4.818pt}{0.400pt}}
\put(220.0,368.0){\rule[-0.200pt]{4.818pt}{0.400pt}}
\put(198,368){\makebox(0,0)[r]{6}}
\put(1416.0,368.0){\rule[-0.200pt]{4.818pt}{0.400pt}}
\put(220.0,495.0){\rule[-0.200pt]{4.818pt}{0.400pt}}
\put(198,495){\makebox(0,0)[r]{8}}
\put(1416.0,495.0){\rule[-0.200pt]{4.818pt}{0.400pt}}
\put(220.0,622.0){\rule[-0.200pt]{4.818pt}{0.400pt}}
\put(198,622){\makebox(0,0)[r]{10}}
\put(1416.0,622.0){\rule[-0.200pt]{4.818pt}{0.400pt}}
\put(220.0,750.0){\rule[-0.200pt]{4.818pt}{0.400pt}}
\put(198,750){\makebox(0,0)[r]{12}}
\put(1416.0,750.0){\rule[-0.200pt]{4.818pt}{0.400pt}}
\put(220.0,877.0){\rule[-0.200pt]{4.818pt}{0.400pt}}
\put(198,877){\makebox(0,0)[r]{14}}
\put(1416.0,877.0){\rule[-0.200pt]{4.818pt}{0.400pt}}
\put(220.0,113.0){\rule[-0.200pt]{0.400pt}{4.818pt}}
\put(220,68){\makebox(0,0){600}}
\put(220.0,857.0){\rule[-0.200pt]{0.400pt}{4.818pt}}
\put(394.0,113.0){\rule[-0.200pt]{0.400pt}{4.818pt}}
\put(394,68){\makebox(0,0){800}}
\put(394.0,857.0){\rule[-0.200pt]{0.400pt}{4.818pt}}
\put(567.0,113.0){\rule[-0.200pt]{0.400pt}{4.818pt}}
\put(567,68){\makebox(0,0){1000}}
\put(567.0,857.0){\rule[-0.200pt]{0.400pt}{4.818pt}}
\put(741.0,113.0){\rule[-0.200pt]{0.400pt}{4.818pt}}
\put(741,68){\makebox(0,0){1200}}
\put(741.0,857.0){\rule[-0.200pt]{0.400pt}{4.818pt}}
\put(915.0,113.0){\rule[-0.200pt]{0.400pt}{4.818pt}}
\put(915,68){\makebox(0,0){1400}}
\put(915.0,857.0){\rule[-0.200pt]{0.400pt}{4.818pt}}
\put(1089.0,113.0){\rule[-0.200pt]{0.400pt}{4.818pt}}
\put(1089,68){\makebox(0,0){1600}}
\put(1089.0,857.0){\rule[-0.200pt]{0.400pt}{4.818pt}}
\put(1262.0,113.0){\rule[-0.200pt]{0.400pt}{4.818pt}}
\put(1262,68){\makebox(0,0){1800}}
\put(1262.0,857.0){\rule[-0.200pt]{0.400pt}{4.818pt}}
\put(1436.0,113.0){\rule[-0.200pt]{0.400pt}{4.818pt}}
\put(1436,68){\makebox(0,0){2000}}
\put(1436.0,857.0){\rule[-0.200pt]{0.400pt}{4.818pt}}
\put(220.0,113.0){\rule[-0.200pt]{292.934pt}{0.400pt}}
\put(1436.0,113.0){\rule[-0.200pt]{0.400pt}{184.048pt}}
\put(220.0,877.0){\rule[-0.200pt]{292.934pt}{0.400pt}}
\put(45,495){\makebox(0,0){$\sigma$ (pb)}}
\put(828,23){\makebox(0,0){$M_{Z'}$ (GeV)}}
\put(220.0,113.0){\rule[-0.200pt]{0.400pt}{184.048pt}}
\put(220,864){\usebox{\plotpoint}}
\multiput(220.58,857.76)(0.499,-1.757){171}{\rule{0.120pt}{1.502pt}}
\multiput(219.17,860.88)(87.000,-301.882){2}{\rule{0.400pt}{0.751pt}}
\multiput(307.58,555.91)(0.499,-0.805){171}{\rule{0.120pt}{0.744pt}}
\multiput(306.17,557.46)(87.000,-138.456){2}{\rule{0.400pt}{0.372pt}}
\multiput(394.00,417.92)(0.778,-0.499){109}{\rule{0.721pt}{0.120pt}}
\multiput(394.00,418.17)(85.503,-56.000){2}{\rule{0.361pt}{0.400pt}}
\multiput(481.00,361.92)(1.669,-0.497){49}{\rule{1.423pt}{0.120pt}}
\multiput(481.00,362.17)(83.046,-26.000){2}{\rule{0.712pt}{0.400pt}}
\multiput(567.00,335.92)(3.727,-0.492){21}{\rule{3.000pt}{0.119pt}}
\multiput(567.00,336.17)(80.773,-12.000){2}{\rule{1.500pt}{0.400pt}}
\multiput(654.00,323.93)(9.616,-0.477){7}{\rule{7.060pt}{0.115pt}}
\multiput(654.00,324.17)(72.347,-5.000){2}{\rule{3.530pt}{0.400pt}}
\multiput(741.00,318.94)(12.618,-0.468){5}{\rule{8.800pt}{0.113pt}}
\multiput(741.00,319.17)(68.735,-4.000){2}{\rule{4.400pt}{0.400pt}}
\put(828,314.17){\rule{17.500pt}{0.400pt}}
\multiput(828.00,315.17)(50.678,-2.000){2}{\rule{8.750pt}{0.400pt}}
\put(915,312.67){\rule{20.958pt}{0.400pt}}
\multiput(915.00,313.17)(43.500,-1.000){2}{\rule{10.479pt}{0.400pt}}
\put(1002,311.67){\rule{20.958pt}{0.400pt}}
\multiput(1002.00,312.17)(43.500,-1.000){2}{\rule{10.479pt}{0.400pt}}
\put(1089,310.67){\rule{20.717pt}{0.400pt}}
\multiput(1089.00,311.17)(43.000,-1.000){2}{\rule{10.359pt}{0.400pt}}
\put(1262,309.67){\rule{20.958pt}{0.400pt}}
\multiput(1262.00,310.17)(43.500,-1.000){2}{\rule{10.479pt}{0.400pt}}
\put(1175.0,311.0){\rule[-0.200pt]{20.958pt}{0.400pt}}
\put(1349.0,310.0){\rule[-0.200pt]{20.958pt}{0.400pt}}
\put(220,329){\usebox{\plotpoint}}
\put(220.0,329.0){\rule[-0.200pt]{292.934pt}{0.400pt}}
\put(220,495){\usebox{\plotpoint}}
\put(220.0,495.0){\rule[-0.200pt]{292.934pt}{0.400pt}}
\end{picture}

\noindent
{\bf Fig. 3:} Top cross section including the Standard Model
and the $Z'$ contribution as a function of the $Z'$ mass. The
two horizontal lines show the upper and lower limits of the current
experimental result obtained by combining CDF and \D0 data \cite{cdf}. 

\bigskip
% GNUPLOT: LaTeX picture
\setlength{\unitlength}{0.240900pt}
\ifx\plotpoint\undefined\newsavebox{\plotpoint}\fi
\sbox{\plotpoint}{\rule[-0.200pt]{0.400pt}{0.400pt}}%
\begin{picture}(1500,900)(0,0)
\font\gnuplot=cmr10 at 10pt
\gnuplot
\sbox{\plotpoint}{\rule[-0.200pt]{0.400pt}{0.400pt}}%
\put(220.0,113.0){\rule[-0.200pt]{292.934pt}{0.400pt}}
\put(220.0,113.0){\rule[-0.200pt]{4.818pt}{0.400pt}}
\put(198,113){\makebox(0,0)[r]{0}}
\put(1416.0,113.0){\rule[-0.200pt]{4.818pt}{0.400pt}}
\put(220.0,198.0){\rule[-0.200pt]{4.818pt}{0.400pt}}
\put(198,198){\makebox(0,0)[r]{1}}
\put(1416.0,198.0){\rule[-0.200pt]{4.818pt}{0.400pt}}
\put(220.0,283.0){\rule[-0.200pt]{4.818pt}{0.400pt}}
\put(198,283){\makebox(0,0)[r]{2}}
\put(1416.0,283.0){\rule[-0.200pt]{4.818pt}{0.400pt}}
\put(220.0,368.0){\rule[-0.200pt]{4.818pt}{0.400pt}}
\put(198,368){\makebox(0,0)[r]{3}}
\put(1416.0,368.0){\rule[-0.200pt]{4.818pt}{0.400pt}}
\put(220.0,453.0){\rule[-0.200pt]{4.818pt}{0.400pt}}
\put(198,453){\makebox(0,0)[r]{4}}
\put(1416.0,453.0){\rule[-0.200pt]{4.818pt}{0.400pt}}
\put(220.0,537.0){\rule[-0.200pt]{4.818pt}{0.400pt}}
\put(198,537){\makebox(0,0)[r]{5}}
\put(1416.0,537.0){\rule[-0.200pt]{4.818pt}{0.400pt}}
\put(220.0,622.0){\rule[-0.200pt]{4.818pt}{0.400pt}}
\put(198,622){\makebox(0,0)[r]{6}}
\put(1416.0,622.0){\rule[-0.200pt]{4.818pt}{0.400pt}}
\put(220.0,707.0){\rule[-0.200pt]{4.818pt}{0.400pt}}
\put(198,707){\makebox(0,0)[r]{7}}
\put(1416.0,707.0){\rule[-0.200pt]{4.818pt}{0.400pt}}
\put(220.0,792.0){\rule[-0.200pt]{4.818pt}{0.400pt}}
\put(198,792){\makebox(0,0)[r]{8}}
\put(1416.0,792.0){\rule[-0.200pt]{4.818pt}{0.400pt}}
\put(220.0,877.0){\rule[-0.200pt]{4.818pt}{0.400pt}}
\put(198,877){\makebox(0,0)[r]{9}}
\put(1416.0,877.0){\rule[-0.200pt]{4.818pt}{0.400pt}}
\put(220.0,113.0){\rule[-0.200pt]{0.400pt}{4.818pt}}
\put(220,68){\makebox(0,0){160}}
\put(220.0,857.0){\rule[-0.200pt]{0.400pt}{4.818pt}}
\put(394.0,113.0){\rule[-0.200pt]{0.400pt}{4.818pt}}
\put(394,68){\makebox(0,0){180}}
\put(394.0,857.0){\rule[-0.200pt]{0.400pt}{4.818pt}}
\put(567.0,113.0){\rule[-0.200pt]{0.400pt}{4.818pt}}
\put(567,68){\makebox(0,0){200}}
\put(567.0,857.0){\rule[-0.200pt]{0.400pt}{4.818pt}}
\put(741.0,113.0){\rule[-0.200pt]{0.400pt}{4.818pt}}
\put(741,68){\makebox(0,0){220}}
\put(741.0,857.0){\rule[-0.200pt]{0.400pt}{4.818pt}}
\put(915.0,113.0){\rule[-0.200pt]{0.400pt}{4.818pt}}
\put(915,68){\makebox(0,0){240}}
\put(915.0,857.0){\rule[-0.200pt]{0.400pt}{4.818pt}}
\put(1089.0,113.0){\rule[-0.200pt]{0.400pt}{4.818pt}}
\put(1089,68){\makebox(0,0){260}}
\put(1089.0,857.0){\rule[-0.200pt]{0.400pt}{4.818pt}}
\put(1262.0,113.0){\rule[-0.200pt]{0.400pt}{4.818pt}}
\put(1262,68){\makebox(0,0){280}}
\put(1262.0,857.0){\rule[-0.200pt]{0.400pt}{4.818pt}}
\put(1436.0,113.0){\rule[-0.200pt]{0.400pt}{4.818pt}}
\put(1436,68){\makebox(0,0){300}}
\put(1436.0,857.0){\rule[-0.200pt]{0.400pt}{4.818pt}}
\put(220.0,113.0){\rule[-0.200pt]{292.934pt}{0.400pt}}
\put(1436.0,113.0){\rule[-0.200pt]{0.400pt}{184.048pt}}
\put(220.0,877.0){\rule[-0.200pt]{292.934pt}{0.400pt}}
\put(45,495){\makebox(0,0){$\sigma(pb)$}}
\put(828,23){\makebox(0,0){$M_D$ (GeV)}}
\put(220.0,113.0){\rule[-0.200pt]{0.400pt}{184.048pt}}
\put(220,815){\usebox{\plotpoint}}
\multiput(220.58,811.61)(0.499,-0.898){171}{\rule{0.120pt}{0.817pt}}
\multiput(219.17,813.30)(87.000,-154.304){2}{\rule{0.400pt}{0.409pt}}
\multiput(307.58,656.05)(0.499,-0.765){171}{\rule{0.120pt}{0.711pt}}
\multiput(306.17,657.52)(87.000,-131.523){2}{\rule{0.400pt}{0.356pt}}
\multiput(394.58,523.91)(0.499,-0.505){171}{\rule{0.120pt}{0.505pt}}
\multiput(393.17,524.95)(87.000,-86.953){2}{\rule{0.400pt}{0.252pt}}
\multiput(481.00,436.92)(0.551,-0.499){153}{\rule{0.541pt}{0.120pt}}
\multiput(481.00,437.17)(84.877,-78.000){2}{\rule{0.271pt}{0.400pt}}
\multiput(567.00,358.92)(0.872,-0.498){97}{\rule{0.796pt}{0.120pt}}
\multiput(567.00,359.17)(85.348,-50.000){2}{\rule{0.398pt}{0.400pt}}
\multiput(654.00,308.92)(1.287,-0.498){65}{\rule{1.124pt}{0.120pt}}
\multiput(654.00,309.17)(84.668,-34.000){2}{\rule{0.562pt}{0.400pt}}
\multiput(741.00,274.92)(1.412,-0.497){59}{\rule{1.223pt}{0.120pt}}
\multiput(741.00,275.17)(84.462,-31.000){2}{\rule{0.611pt}{0.400pt}}
\multiput(828.00,243.92)(1.566,-0.497){53}{\rule{1.343pt}{0.120pt}}
\multiput(828.00,244.17)(84.213,-28.000){2}{\rule{0.671pt}{0.400pt}}
\multiput(915.00,215.92)(3.727,-0.492){21}{\rule{3.000pt}{0.119pt}}
\multiput(915.00,216.17)(80.773,-12.000){2}{\rule{1.500pt}{0.400pt}}
\multiput(1002.00,203.92)(2.960,-0.494){27}{\rule{2.420pt}{0.119pt}}
\multiput(1002.00,204.17)(81.977,-15.000){2}{\rule{1.210pt}{0.400pt}}
\multiput(1089.00,188.92)(2.926,-0.494){27}{\rule{2.393pt}{0.119pt}}
\multiput(1089.00,189.17)(81.033,-15.000){2}{\rule{1.197pt}{0.400pt}}
\multiput(1175.00,173.94)(12.618,-0.468){5}{\rule{8.800pt}{0.113pt}}
\multiput(1175.00,174.17)(68.735,-4.000){2}{\rule{4.400pt}{0.400pt}}
\multiput(1262.00,169.92)(4.081,-0.492){19}{\rule{3.264pt}{0.118pt}}
\multiput(1262.00,170.17)(80.226,-11.000){2}{\rule{1.632pt}{0.400pt}}
\put(1349,158.67){\rule{20.958pt}{0.400pt}}
\multiput(1349.00,159.17)(43.500,-1.000){2}{\rule{10.479pt}{0.400pt}}
\put(220,738){\usebox{\plotpoint}}
\multiput(220.58,734.68)(0.499,-0.875){171}{\rule{0.120pt}{0.799pt}}
\multiput(219.17,736.34)(87.000,-150.342){2}{\rule{0.400pt}{0.399pt}}
\multiput(307.58,583.07)(0.499,-0.759){171}{\rule{0.120pt}{0.707pt}}
\multiput(306.17,584.53)(87.000,-130.533){2}{\rule{0.400pt}{0.353pt}}
\multiput(394.00,452.92)(0.530,-0.499){161}{\rule{0.524pt}{0.120pt}}
\multiput(394.00,453.17)(85.912,-82.000){2}{\rule{0.262pt}{0.400pt}}
\multiput(481.00,370.92)(0.632,-0.499){133}{\rule{0.606pt}{0.120pt}}
\multiput(481.00,371.17)(84.742,-68.000){2}{\rule{0.303pt}{0.400pt}}
\multiput(567.00,302.92)(0.890,-0.498){95}{\rule{0.810pt}{0.120pt}}
\multiput(567.00,303.17)(85.318,-49.000){2}{\rule{0.405pt}{0.400pt}}
\multiput(654.00,253.92)(1.511,-0.497){55}{\rule{1.300pt}{0.120pt}}
\multiput(654.00,254.17)(84.302,-29.000){2}{\rule{0.650pt}{0.400pt}}
\multiput(741.00,224.92)(1.460,-0.497){57}{\rule{1.260pt}{0.120pt}}
\multiput(741.00,225.17)(84.385,-30.000){2}{\rule{0.630pt}{0.400pt}}
\multiput(828.00,194.92)(2.098,-0.496){39}{\rule{1.757pt}{0.119pt}}
\multiput(828.00,195.17)(83.353,-21.000){2}{\rule{0.879pt}{0.400pt}}
\multiput(915.00,173.92)(3.431,-0.493){23}{\rule{2.777pt}{0.119pt}}
\multiput(915.00,174.17)(81.236,-13.000){2}{\rule{1.388pt}{0.400pt}}
\multiput(1002.00,160.92)(3.431,-0.493){23}{\rule{2.777pt}{0.119pt}}
\multiput(1002.00,161.17)(81.236,-13.000){2}{\rule{1.388pt}{0.400pt}}
\multiput(1089.00,147.93)(6.519,-0.485){11}{\rule{5.014pt}{0.117pt}}
\multiput(1089.00,148.17)(75.593,-7.000){2}{\rule{2.507pt}{0.400pt}}
\multiput(1175.00,140.93)(6.595,-0.485){11}{\rule{5.071pt}{0.117pt}}
\multiput(1175.00,141.17)(76.474,-7.000){2}{\rule{2.536pt}{0.400pt}}
\multiput(1262.00,133.93)(9.616,-0.477){7}{\rule{7.060pt}{0.115pt}}
\multiput(1262.00,134.17)(72.347,-5.000){2}{\rule{3.530pt}{0.400pt}}
\multiput(1349.00,128.94)(12.618,-0.468){5}{\rule{8.800pt}{0.113pt}}
\multiput(1349.00,129.17)(68.735,-4.000){2}{\rule{4.400pt}{0.400pt}}
\end{picture}

\noindent
{\bf Fig. 4} The $D$ exotic quark cross section as a function of its mass.
The lower curve corresponds to pure QCD production and the upper curve
includes QCD plus the $Z'$ contribution with $M_{Z'}=1$ TeV. 

\newpage

\psfig{file=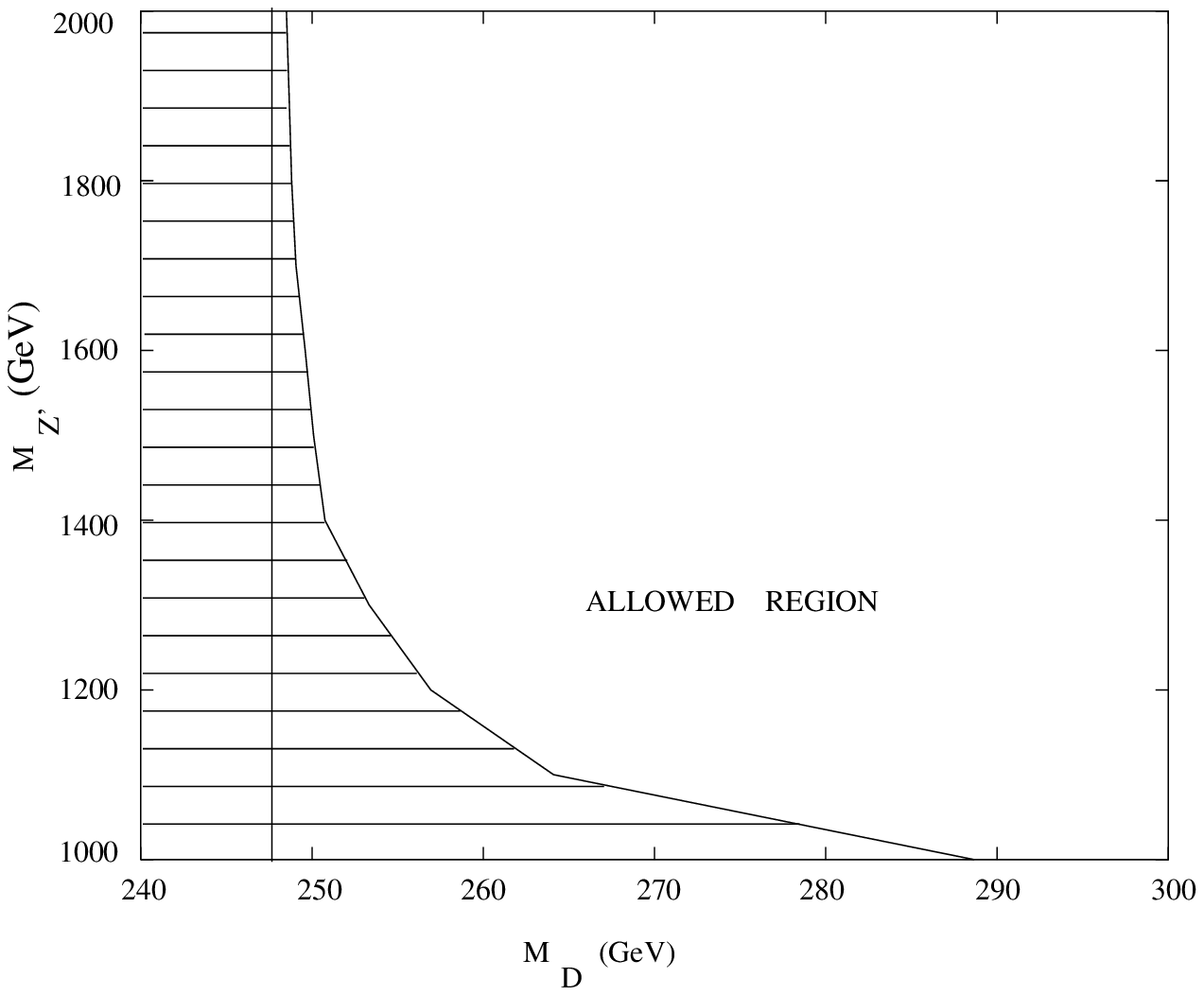}

\noindent
{\bf Fig. 5} The lower bound on the exotic quark as a function of its
mass $M_D$ and the $Z'$ mass. The verticle line corresponds to the 
limit obtained by excluding the $Z'$ contribution to the $D\bar D$
cross section. 

\bigskip
\noindent
{\bf Acknowledgements:} We thank Phil Baringer for discussions about 
backgrounds and Marc Paterno and Sarah Eno for communications that 
patiently explained the cuts and efficiencies in \cite{susy}. 
Computational facilities of the Kansas Institute for Theoretical Science
were used in this work, which was supported in part by U.S. DOE Grant No. 
DE-FG02-85ER40214.


\begin{thebibliography}{}
\bibitem{cdf} We show the combined CDF and \D0 number $\sigma_{tt}=6.7\pm 1.3
pb$ (preliminary). See G. Velev ``CDF Findings of the Top Quark", and B. Klina
``\D0 Findings on the Top Quark", available on the web pages of the
collaborations.
\end{thebibliography}
\end{document}